\documentclass{elsart}
\usepackage{graphicx}
\usepackage{amssymb}
\begin{document}
\begin{frontmatter}
\title{Bipartite Producer-Consumer Networks and the Size Distribution of Firms}
\author{Wang Dahui,}
\author{Zhou Li,}
\author{Di Zengru\corauthref{cor1}}


\corauth[cor1]{Corresponding author.
\\ {Tel.: +86-10-58807876; fax:+86-10-58807876.}
\\{\em E-mail address:\/}\, zdi@bnu.edu.cn}
\address{Department of Systems Science, School of Management, Beijing Normal University, Beijing,
100875, China
}
\address{Center for Complexity Research,Beijing Normal
University,Beijing,100875,China}
\begin{abstract}

A bipartite producer-consumer network is constructed to describe
the industrial structure. The edges from consumer to producer
represent the choices of the consumer for the final products and
the degree of producer can represent its market share. So the size
distribution of firms can be characterized by producer's degree
distribution. The probability for a producer receiving a new
consumption is determined by its competency described by initial
attractiveness and the self-reinforcing mechanism in the
competition described by preferential attachment. The cases with
constant total consumption and with growing market are studied.
The following results are obtained: 1, Without market growth and a
uniform initial attractiveness $a$, the final distribution of firm
sizes is Gamma distribution for $a>1$ and is exponential for
$a=1$. If $a<1$, the distribution is power in small size and
exponential in upper tail; 2, For a growing market, the size
distribution of firms obeys the power law. The exponent is
affected by the market growth and the initial attractiveness of
the firms.
\end{abstract}

\begin{keyword}size distribution of firms \sep bipartite networks \sep complex networks \sep econophysics
\\
\PACS 89.65.Gh, 89.75.Da, 89.75.Hc

\end{keyword}

\end{frontmatter}
\section{Introduction}

Industrial structure is an important issue both in macroeconomic
and physical investigations. It is closely related to the dynamics
of firms and market. From the empirical studies, it has been found
that there are several ``stylized facts" related to the processes
of industrial evolution. One of them is the skewed distribution of
firm size which is usually measured by the sales, the number of
employees, the capital employed and the total assets. Such skewed
distribution has usually been described by lognormal distribution
since Gibrat\cite{Gibratlaw} and its upper tail has been described
by Pareto distribution or Zipf's law\cite{SimonSkew,Lucas}. In
terms of cumulative distribution $P_{>}(x)$ for firm size $x$,
this states that $P_{>}(x)\varpropto x^\mu$ for larger $x$, where
$\mu$ is the exponent called Pareto index. Recently, there are
more empirical researches investigating the properties of firm
size distribution in detail\cite{MStanley95,RamsdenEmp}. Axtell
reveals that the U.S. firm size is precisely described by power
law distribution\cite{AxtellSic}. For the cumulative distribution
of firm sizes by employees, the index is 1.059, and for the
cumulative distribution of firm sizes by receipts in dollars the
index is 0.994. Some scholars in Italy and Japan, by exploring the
size distribution of European and Japanese firms in detail, have
found more evidence for the power law distribution of firm size in
upper tail including the Zipf law in firms bankruptcy or
extinction\cite{ItalyEm,JapanEm}. All the indexes for cumulative
distributions are ranging around 1 (from 0.7 to 1.2).

Various kinds of power-law behaviors have been observed in a wide
range of systems, including the wealth distribution of
individuals\cite{Money,Market,Wang} and the price-returns in stock
markets\cite{Econophysics,Return}. Pareto-Zipf law in firm size
provides another interesting example which exhibits some universal
characteristics similar to those observed in physical systems with
a large number of interacting units. Hence the growth dynamics and
size distribution of business firms have become a subject of
interest among economists and physicists, especially those who
working in econophysics\cite{Econophysics}. Together with the
works in macroeconomics\cite{EcoSize1,EcoSize2,EcoSize3}, many
efforts have been done from the perspectives of physics in
accordance with these empirical facts. Takayasu advanced an
aggregation-annihilation reaction model to study firm size
dynamics\cite{Takayasu}. Axtell has argued that complexity
approach should be used to deal with the problem, and agent based
modelling together with evolutionary dynamics should be helpful to
understand the formation of power law\cite{AxtellModel}. Amaral et
al. have studied firm growth dynamics since 1997\cite{Amaral97}.
They have developed a stochastic model based on interactions
between different units of a complex system. Each unit has a
complex internal structure comprising many subunits and its growth
dynamics is dependent on the interactions between them. This
model's prediction goes well with the empirical
result\cite{AmaralPRL12,AmaralPhyA}. Some other models have also
been presented, which are based on the competition
dynamics\cite{GuptaModel}, the information transition, herd
behaviors\cite{ZhengModel}, and the proportional growth for the
firms' sizes and the number of independent constituent units of
the firms.\cite{FabratSize}.

The development of the research on complex
networks\cite{Watts,AlbertNature,AlbertStat} has given us a new
perspective to speculate the power law distribution of firm size.
First, it provides us a universal tool for the research of complex
systems\cite{AmaralNet}. Actually, any complex systems made up by
the interactive components can be described by networks, in which
the components are represented by the vertices, and the
interactions by the edges. Second, the empirical results
demonstrate that many large networks in the real world are scale
free, like the World Wide Web, the internet, the network of
movie-actor collaborations, the network of citations of scientific
papers, the network of scientific collaborations and so on. They
all have a scale-free degree distribution with tails that decay as
a power law (see \cite{AlbertStat,AmaralNet} as reviews). So the
complex networks in the nature give us  examples of power law
behavior. Barab\'{a}si and Albert have argued that the scale-free
nature of real networks is rooted in two generic mechanisms, i.e.
the growth and preferential attachment\cite{BaraGP}. We hope the
mechanism responsible for the emergence of scale-free networks
would give clues to understand the power law distribution of firm
size. Actually the network approach has been already applied to
economic analysis. Souma et al. have done some empirical studies
on business networks. The results reveal the possibility that
business networks will fall into the scale-free
category\cite{SoumaNet}. Garlaschelli and Loffredo have argued
that the outcome of wealth dynamics depends strongly on the
topological properties of the underlying transaction
network\cite{WealthinNet}. The topology and economic cycle
synchronization on the world trade web have also been
studied\cite{Li}.

Actually, the dynamics of firms and market could also be precisely
described by the network approach. We can consider that producers
and consumers are the two kinds of vertices and they are related
with each other by links. So they could be represented as a
bipartite network, which includes two kinds of vertices and whose
edges only connect vertices of different kinds. The edges between
producers and consumers can stand for the consumers' choices for
their consumptions. To explore the size distribution of firms, we
assume that every edge stands for one unit of consumption and the
degree of a producer describes its sales. Then the size
distribution of firms is corresponding to the degree distribution
of producers. As the results of market competition, consumers can
change their choices of the consumption, which refers to the
switches of links between producers. The mechanism of preferential
attachment is just a good description for the rich-getting-richer
effect or the self-reinforcing mechanism in the market
competition. So it is a natural way to study the formation of size
distribution of firms by investigating the evolution of the
network.

Bipartite network is an important kind of networks in real world
and the collaboration networks of movie
actors\cite{Watts,WattsSmall} and scientists\cite{NewmanSci} are
the typical ones. The bipartite producer-consumer network we
discuss here has different properties compared with the above
collaboration networks. The links between collaborators and acts
in collaboration networks are fixed but it can also be rewired
with the evolving of the network in the producer-consumer network
model. So the study of producer-consumer network is also valuable
to understand the properties of this kind of bipartite networks.

The presentation is organized as following. In section 2, the
model $A$ with the constant total consumption is discussed, in
which producers compete in a constant market. The results reveal
that there is no power law distribution of firm sizes in upper
tail. In section 3 we investigate the more realistic case with
growing markets. In model $B$, the number of producers and the
market, which is described by total consumptions, both grow with
the time. Led by the mechanism of preferential attachment, the
size distribution of firms obeys the power law, and the exponent
is affected by the growth and the initial attractiveness of the
firms. In section 4, we summarize our results and give concluding
remarks.

\section{Model $A$: network evolving with constant total consumption }

In the industrial structure, the scale effect of the firms
determined by their technological levels is one of the factors
that influence the firm size. Another one is the self-reinforcing
effect in market competition. Assume that there are $N$ producers
and $K$ consumers in the market. They form a bipartite network and
the edges connect the consumers to producers. For simplicity and
without losing any generality, we assume one consumer has only one
edge which represents one unit of consumption. The degree of
producer describes its size by means of market share, so the size
distribution of firms is characterized by the degree distribution
of producers. In model $A$, we consider the situation with
constant total consumption $K$. The total number of edges will not
been changed, but as the results of competition, the consumer
could switch between producers, which means some edges will be
rewired. Then we concentrate on the final steady degree
distribution of the model. Let $N_{k}(t)$ denotes the number of
vertices with degree $k$ at time step $t$. From any given initial
distribution, the model evolves as following two steps:

1, Cutting one edge randomly at each time step. Let $n_{k}(t)$
indicate the number of vertices with degree $k$ after cutting one
edge at randomly. It is determined by

\begin{equation}
n_{k}(t)=\frac{k+1}{K}N_{k+1}(t-1)+(1-\frac{k}{K})N_{k}(t-1)\label{n}
\end{equation}

2, Connecting the edge to the producer with preferential
attachment mechanism. The probability of connecting the edge to
one producer with degree $k$ is: $\frac{k+a}{K-1+Na}$, where $a$
is a parameter called the initial attractiveness of the node. That
is related to the intrinsic competence of the firm in our
discussion. It is including the technology, the distinctions of
the product, and other initial features of the firm. Due to the
diversity of the demand, we assume any firm has the same initial
attractiveness without losing any generality. Hence, the number of
vertices with degree $k$ after rewired is:

\begin{equation}
N_{k}(t)=\frac{k-1+a}{K-1+Na}n_{k-1}(t)+(1-\frac{k+a}{K-1+Na})n_{k}(t)\label{N}
\end{equation}

The eqs.(\ref{n}) and (\ref{N}) give the dynamics of network
evolution. The boundary conditions are
\begin{eqnarray*}
n_{0}(t)&=&\frac{N_{1}(t-1)}{K}+N_{0}(t-1)\\
N_{0}(t)&=&(1-\frac{a}{K-1+Na})n_{0}(t)\\
n_{K}(t)&=&0\\
N_{K}(t)&=&\frac{K-1+a}{K-1+Na}n_{K-1}(t)
\end{eqnarray*}
The eqs.(\ref{n}) and (\ref{N}) with the above boundary conditions
give:
\begin{eqnarray*}
&&\sum_{k=0}^{k=K}n_{k}(t)=\sum_{k=0}^{k=K}N_{k}(t)=\sum_{k=0}^{k=K}N_{k}(t-1)=N\\
&&\sum_{k=0}^{k=K}kn_{k}(t)=\sum_{k=0}^{k=K}kN_{k}(t-1)-1=K-1\\
&&\sum_{k=0}^{k=K}kN_{k}(t)=\sum_{k=0}^{k=K}kn_{k}(t)+1=K
\end{eqnarray*}

These results approve that we have indeed cut one edge in
eq.(\ref{n}) and rewired it in eq.(\ref{N}), while the total
number of producers and consumers are all constant.

\begin{figure}
\centering{\includegraphics[width=7cm]{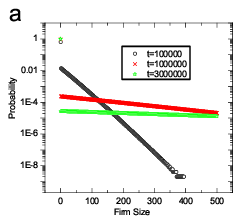}}
\includegraphics[width=7.2cm]{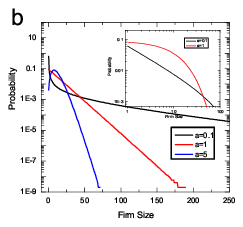}\includegraphics[width=8cm]{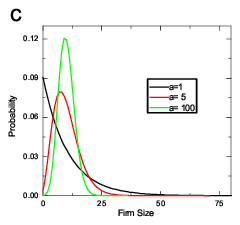}
\caption{The stationary probability distribution of firm size
without market growth. (a) The results for $a=0$. Many producers
fail in the competition. Almost all consumers connect with few
producers. (b) The results for $a\neq0$. In the case of $a=1$ and
for the upper tail in the case of $a<1$, the firm size obeys the
exponential distribution. When $a>1$, the firm size obeys Gamma
distribution. The upper inset indicates that the lower tail of
firm size distribution obeys the power law in the case of $a<1$.
(c)The firm size distribution in the case of $a=1$, $a>1$ in the
linear coordinates.} \label{cluster}
\end{figure}

Now we can obtain the stable distribution of firm sizes by
numerical solutions of the rate equations (\ref{n}) and (\ref{N})
with boundary conditions and we have obtained the numerical
solutions for the system with total $N=500$ producers and $K=5000$
consumers. From any given initial distributions, the system's
stable distribution is discovered to be related with the
parameters. The simulation results are shown in Figure 1. When
$a=0$, all the producers have no initial attractiveness, i.e.
their advantages in competition are all from the self-reinforcing
mechanism. The results indicate that many producers fail in the
competition. Almost all consumers connect with few producers. In
the case of $a=1$ and for the upper tail in the case of $a<1$, the
firm size obeys the exponential distribution. When $a>1$, the firm
size obeys Gamma distribution. These results are similar with
money distribution in ref \cite{Money,Market} gained by
transferring model. The above results indicate that we can not get
power law distribution by preferential attachment in a constant
market. We will discuss the case of the growing market in the next
section.

\section{Model $B$: network evolving with growing market}

Model $A$ describes the firm size distribution in a constant
market whose growth rate is zero. Due to the technical progress
and the enlargement of inputs, averagely, the total demand and
supply always grow with the time. So we set up another model to
depicts them as following. At every time step, one new producer
and $l$ new consumers enter the system. The new consumers connect
existing producers with preferential probability. Meanwhile, one
old consumer in the system could still switch between different
producers. In contrast to model $A$, at each evolution step, the
number of producers will increase by one and the number of
consumers by $l$.

Supposing that we have $\tilde{K}$ consumers randomly distributed
in $\tilde{N}$ producers in the initial. There are $l$ consumers
and one producer enter the system at every time step from $t=0$.
So at time $t$, we have $K=\tilde{K}+lt$ edges and $N=\tilde{N}+t$
producers. Let $N_{k}(t)$ denotes the number of vertices with
degree $k$ at time step $t$. The network evolves as following:

1, Cutting one edge randomly at each time step. Then the number of
vertices with degree $k$ ($n_{k}(t)$) is given by
\begin{equation}
n_{k}(t)=\frac{k+1}{K}N_{k+1}(t-1)+(1-\frac{k}{K})N_{k}(t-1)\label{eq.7}
\end{equation}

2, Connecting the one old edge and $l$ new links to producer with
preferential attachment mechanism. We get
\begin{equation}
N_{k}(t)=n_{k}(t)+\frac{(l+1)(k-1+a)n_{k-1}(t)}{K-1+Na}-\frac{(l+1)(k+a)n_{k}(t)}{K-1+Na}\label{eq.8}
\end{equation}

3, Adding one new producer to the market. So the boundary
conditions are
\begin{eqnarray*}
&&n_{0}(t)=N_{0}(t-1)+\frac{1}{K}N_{1}(t-1)\\
&&N_{0}(t)=n_{0}(t)-\frac{(l+1)an_{0}(t)}{K-1+Na}+1
\end{eqnarray*}

Combine eq.(\ref{eq.7}), eq.(\ref{eq.8}) and the boundary
conditions, we have:
\begin{eqnarray*}
&&\sum_{k=0}n_{k}(t)=\sum_{k=0}N_{k}(t-1)\\
&&\sum_{k=0}N_{k}(t)=\sum_{k=0}n_{k}(t)+1=\sum_{k=0}N_{k}(t-1)+1\\
&&\sum_{k=0}kn_{k}(t)=\sum_{k=0}kN_{k}(t-1)-1\\
&&\sum_{k=0}kN_{k}(t)=\sum_{k=0}kn_{k}(t)+l+1
\end{eqnarray*}
These equations give the time evolution of the system of $l$
consumers with one producer entering the system at every time
step.

\begin{figure}
\includegraphics[width=7.2cm]{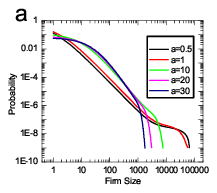}\includegraphics[width=7cm]{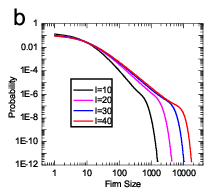}
\caption{The stationary probability distribution of firm size with
market growth. The distribution obeys power law. (a)The effects of
initial attractiveness on the firm size distribution with constant
$l=30$. The exponents range from 1.85 to 2.47. (b)The effects of
$l$ on the firm size distribution. The initial attractiveness
$a=10$. With the increase of $l$, the exponent changes from 2.80
to 2.01.} \label{Fig2}
\end{figure}

We have investigated the properties of this model by numerical
solutions. With $\tilde{K}=100$ consumers randomly distributed
among the $\tilde{N}=100$ producers in the initial, the final
distributions are got by the numerical solutions. If $a=0$, The
results are qualitatively the same as the case of constant market.
That is many producers fail in the competition. Almost all
consumers connect with few producers. The frequency distributions
for $a\neq0$ are shown in Figure 2, which indicate that the size
distribution of firms obeys the power law and the exponent is
related with the market growth $l$ and the initial attractiveness
$a$. Larger $l$ leads to less steep slope and bigger initial
attractiveness leads to steeper slope. The exponents range from
1.85 to 2.80.

\begin{figure}
\centering{\includegraphics[width=9cm]{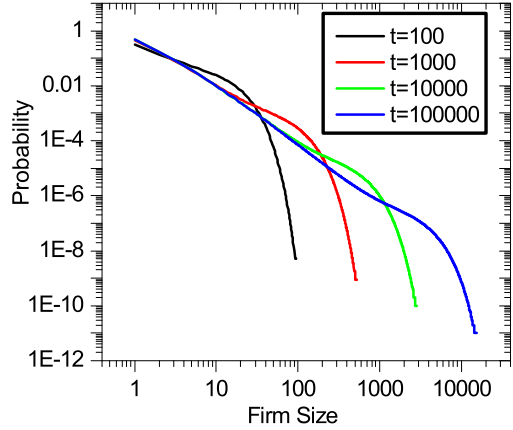}} \caption{The
numerical solutions for different time steps. The exponential tail
moves to upper end with the increase of time steps. $a=10$,
$l=10$.} \label{Fig21}
\end{figure}

From our numerical solutions of the model, we have found that the
exponential tails in the numerical results are due to the limited
runs of the model. If we simulate the model for longer time steps,
the exponential tail will be moved to the upper end (as shown in
Figure \ref{Fig21}). So we believe that when time goes to
infinity, the model with preferential choice of consumption will
result in the power law distribution in the upper tail. The
results of the model are consistent with that of the empirical
studies especially when $a$ is small.

\begin{figure}
\centering{\includegraphics[width=9cm]{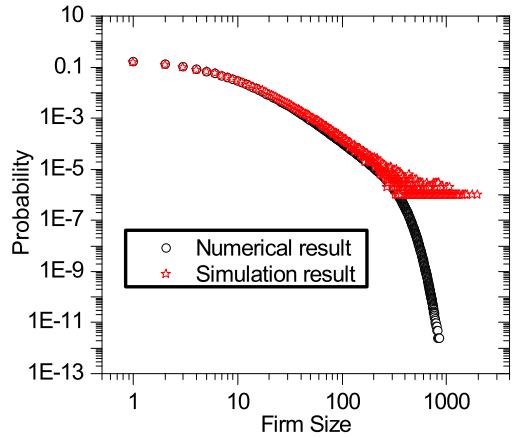}} \caption{The
comparison between the numerical solution of the system and the
simulation result for the case $a=1$, and $l=10$. They consist
well.} \label{Fig3}
\end{figure}

We have done a series of computer simulations for the model $B$
with the same initial conditions. The simulation results are
consistent well with those of numerical solutions of the system.
We show one case in figure 3, in which $a=10$, $l=10$ and the
simulation steps are 1000000.

\section{Summary}

We proposed a bipartite producer-consumer network to investigate
the firm size distribution in this paper. The market dynamics is
described by the evolution of the network and the firm size is
characterized by its market share which is represented by the
degree of producer. The instinct of competition of the firms and
the self-reinforcing mechanism are introduced into the probability
\begin{marginpar}{\rule{1mm}{10mm}}
\end{marginpar}
of connection and the consumers switch their links between the
producers as for competition.  The results indicate that the
economic growth is an important condition for the power law
distribution of firm size distribution. The growth rate and
initial attractiveness of the firms will affect the exponents of
the firm size distribution. Without economic growth, our results
indicate that the initial attractiveness of firms $a$ is an
important parameter to determine the final distribution. The final
distribution is Gamma distribution when $a>1$, is exponential when
$a=1$. When $a<1$, the upper tail is exponential and the lower end
is power. If $a=0$, which means only the self-reinforcing
mechanism works in the market competition, there would be only
fewer producers surviving in the market. All these results provide
understandings to the mechanism of power law distribution of firm
sizes and they maybe valuable for investigating the properties of
bipartite networks.

\section*{Acknowledgments}
Thanks for the referees' helpful comments and suggestions. This
research is supported partially by the National Science Foundation
of China under Grant No.70371072, No.70431002 and No.70471080.


\begin{thebibliography}{99}


\bibitem{Gibratlaw}R. Gibrat, Les InT\`{e}gali\'{e}s Economiques, Sirey, Paris,1931.
\bibitem{SimonSkew}Y. Ijiri, H.A. Simon, Skew Distributions and the Size of Business
Firms, North-Holland, Amsterdam, 1977.
\bibitem{Lucas}R. Lucas, Bell J. Econom.9 (1978) 508.
\bibitem{MStanley95}M.H.R. Stanley, et al., Econom. Lett. 49 (1995) 453-457.
\bibitem{RamsdenEmp}J.J. Ramsden and Gy. Kiss-Hayp¨¢l, Physica A 277, (2000) 220-227. [29]
\bibitem{AxtellSic}R. Axtell, Science 293 (2001) 1819;
\bibitem{ItalyEm}E. Gaffeo, M. Gallegati, A. Palestrini, Physica A 324 (2003)
117-123; C. Di Guilmi; M. Gallegati, P. Ormerodc, Physica A 334
(2004) 267-273.
\bibitem{JapanEm} Y. Fujiwara, C. Di Guilmi, H. Aoyama, M.
Gallegati, W. Souma, Physica A 335 (2004) 197-216; Y. Fujiwara,
Physica A 337 (2004) 219-230; Y. Fujiwara, H. Aoyama, C. Di
Guilmi, W. Souma, M. Gallegati, Physica A 344 (2004) 112-116.
\bibitem{Money}A. Chakraborti and B. K. Chakrabarti, Eur.Phys.J.B 17,(2000) 167-170.
\bibitem{Market}A. Chatterjee et al., Physica A 335 (2004) 155-163.
\bibitem{Wang}N. Ding, Y. Wang, J. Xu, N. Xi, Int. J. of Modern Physics B, 18(17-19) (2004) 2725.
\bibitem{Econophysics} R. N. Mantegna and H. E. Stanley, {\em An Introduction to Econophysics\/},
Cambridge university press, Cambridge, 2000.
\bibitem{Return}P. Gopikrishnan, V. Plerou, L.A.N. Amaral, M.
Meyer, H.E. Stanley, Phys. Rev. E 60 (1999) 5305.
\bibitem{EcoSize1}P. F. Peretto, European Economic Review 43 (1999) 1747-1773.
\bibitem{EcoSize2}F. Hashemi, J. Evol. Econ. (2000) 10: 507-521.
\bibitem{EcoSize3}S. G. Winter, Y. M. Kaniovski, and G. Dosi, J. Evol. Econ. (2003) 13: 355-383.
\bibitem{Takayasu}H.Takayasu and K.Okuyama, Fractals, 6,67-79(1998).
\bibitem{AxtellModel}R. Axtell, CSED Working Paper No.3, Brookings Institution,
2001.
\bibitem{Amaral97}L.A.N. Amaral, S.V. Buldyrev, H. Leschhorn, P. Maass, M. A. Salinger,
H.E. Stanley and M.H.R. Stanley, J. Phys. I France 7, (1997)
521-633; S.V. Buldyrev, L.A.N. Amaral, S. Havlin, H. Leschhorn, P.
Maass, M. A. Salinger, H.E. Stanley and M.H.R. Stanley, J. Phys. I
France 7, (1997) 635-650.
\bibitem{AmaralPRL12}L.A.N. Amaral et al., Phy. Rew. Lett. 80, (1998) 1385; Y.Lee et al., Phys.
Rev. Lett. 81, (1998) 3275.
\bibitem{AmaralPhyA}L.A.N. Amaral et al., Physica A 299 127-136(2001)
\bibitem{GuptaModel}H. M. Gupta and J. R. Campanha, Physica A 323 (2003)
626-634.
\bibitem{ZhengModel}Dafang Zheng; G.J. Rodgersa, P.M. Huic, Physica A 310 (2002)
480-486.
\bibitem{FabratSize}G.De Fabritiis et al., Physica A 324(2003)38-44.
\bibitem{Watts}Watts and Strogatz, Nature 393 (1998) 440.
\bibitem{AlbertNature}R. Albert, H. Jeong, A. L. Barab\'{a}si, Nature 401 (1999) 130.
\bibitem{AlbertStat}R. Albert, A.L.Barab\'{a}si, Rev. Mod. Phys. (2002) 74.
\bibitem{AmaralNet}L.A.N. Amaral, and J.M. Ottino, Eur. Phys. J. B 38, (2004) 147-162.
\bibitem{BaraGP}A.-L. Barab\'{a}si, R. Albert, and H. Jeong, Physica A 272 (1999) 173.
\bibitem{SoumaNet}W. Souma, Y. Fujiwara, H. Aoyama, Physica A 324 (2003) 396-401.
\bibitem{WealthinNet}D. Garlaschelli, M. I. Loffredob, Physica A 338 (2004)
113-118.
\bibitem{Li}X. Li, Y. Jin, G. Chen, Physica A 328 (2003) 287-296.
\bibitem{WattsSmall}D. J. Watts, Small Worlds (1999) (Princeton, New Jersey:
Princeton University Press).
\bibitem{NewmanSci}M.E.J. Newman, Proc. Nat. Acad. Sci. U.S.A. 98, (2001) 404; M. E. J. Newman,
Phys. Rev. E, 64, (2001) 016131; Phys. Rev. E, 64, (2001) 016132.
\end{thebibliography}
\end{document}